\documentclass[runningheads]{llncs}
\usepackage[T1]{fontenc} 
\usepackage{graphicx,verbatim}
\usepackage{amsmath}
\usepackage[ruled,vlined]{algorithm2e} 
\usepackage{algorithmic,float}
\usepackage{tabularx}
\usepackage{booktabs}

\begin{document}

\title{BOLDSimNet: Examining Brain Network Similarity between Task and Resting-State fMRI}

%% Added for anonymized MICCAI 2025 submission

%\author{Anonymized Authors}  
%\authorrunning{Anonymized Author et al.}
%\institute{Anonymized Affiliations \\
%    \email{email@anonymized.com}}
%\maketitle
\author{Boseong Kim\inst{1}, Debashis Das Chakladar\inst{2}, Haejun Chung\inst{1}, and Ikbeom Jang\inst{3}}  %% Added for anonymized MICCAI 2025 submission
\authorrunning{B. Kim et al.}
\institute{
Hanyang University, Seoul 04763, Republic of Korea \and
Luleå University of Technology, 97187 Luleå, Sweden \and
Hankuk University of Foreign Studies, Yongin 17035, Republic of Korea \\
\email{ijang@hufs.ac.kr}
}
\maketitle

\begin{abstract} % word count: 189 / 150~250 limit
Traditional causal connectivity methods in task-based and resting-state functional magnetic resonance imaging (fMRI) face challenges in accurately capturing directed information flow due to their sensitivity to noise and inability to model multivariate dependencies. These limitations hinder the effective comparison of brain networks between cognitive states, making it difficult to analyze network reconfiguration during task and resting states. To address these issues, we propose BOLDSimNet, a novel framework utilizing Multivariate Transfer Entropy (MTE) to measure causal connectivity and network similarity across different cognitive states. 
%Unlike conventional approaches, MTE effectively captures nonlinear, multivariate interactions and provides a more robust measure of directed information flow, enabling precise assessment of connectivity changes. 
Our method groups functionally similar regions of interest (ROIs) rather than spatially adjacent nodes, improving accuracy in network alignment. We applied BOLDSimNet to fMRI data from 40 healthy controls and found that children exhibited higher similarity scores between task and resting states compared to adolescents, indicating reduced variability in attention shifts. In contrast, adolescents showed more differences between task and resting states in the Dorsal Attention Network (DAN) and the Default Mode Network (DMN), reflecting enhanced network adaptability. These findings emphasize developmental variations in the reconfiguration of the causal brain network, showcasing BOLDSimNet’s ability to quantify network similarity and identify attentional fluctuations between different cognitive states.

\keywords{Brain connectivity network \and Graph similarity \and fMRI}
\end{abstract}

\section{Introduction}
When switching between task and resting states, the brain undergoes substantial reorganization of its functional networks \cite{cole2014intrinsic}. A study of resting-state fMRI (rs-fMRI) observed increased activity of the default mode network (DMN) during rest, suggesting its potential as a key marker of functional disorders of the brain \cite{van2010exploring}. In a related task-based fMRI (ts-fMRI) study comparing children and adults, children primarily exhibited hypoactivation in the somatomotor network (SMN) and dorsal attention network (DAN), whereas adults showed increased activation in the DAN and reduced SMN hypoactivation \cite{cortese2012toward}. In addition to identifying isolated activated brain regions, it is crucial to analyze functional connectivity within the brain network \cite{fox2005human}. In an fMRI study using Granger causality (GC), the authors demonstrated that cognitive load is indeed reflected in the strength of causal interactions and presented a method for exploring Region of Interest (ROI)-level causal relationships within the brain network \cite{roebroeck2005mapping}. However, there were limitations that prevented it from fully accounting for the nonlinear nature of fMRI \cite{deneux2006using}. To overcome the issue, transfer entropy (TE) provides a nonlinear method of capturing dynamic, direction-specific information flow between time-series data \cite{wu2021effective}. Multivariate transfer entropy (MTE) demonstrated improved performance in capturing brain connectivity between multiple brain regions that interact \cite{wibral2014transfer}. 

Numerous studies have underscored the significance of recognizing similarities or differences in brain connectivity networks between task-based and resting-state fMRI conditions \cite{zhang2016characterizing,di2013task,mheich2020brain}. However, comparing brain networks of different sizes has proven challenging, as discrepancies in the number of nodes complicate the direct assessment of network similarity \cite{van2010comparing}. In comparing differently sized fMRI brain networks, researchers have used either a fixed threshold to standardize graph size \cite{achard2006resilient} or removed nodes from larger networks to match smaller ones \cite{zalesky2010whole}, yet both methods can disrupt the original network structure—either by losing key connections or keeping unimportant ones \cite{hayasaka2010comparison}.
SimBrainNet, an Electroencephalography (EEG)-based approach for measuring brain networks, leverages the spatial adjacency of EEG channels to compute similarity among different networks efficiently \cite{das2024simbrainnet}. However, fMRI data reveal that physically adjacent cortical regions do not necessarily exhibit functional similarity \cite{yeo2011}, suggesting that fMRI studies should emphasize functional relationships rather than mere anatomical proximity.

% We make two primary contributions. First, the proposed BOLDSimNet method measures the similarity between MTE-based brain networks for two cognitive
% tasks. Second, we introduce a functional imaging-specific approach to substitute, insert, or delete nodes when comparing two brain connectivity graphs. It uses functionally similar ROI groups to calculate similarity while minimizing graph transformation loss.
%  when comparing brain connectivity graphs of different sizes
We make two primary contributions. First, the proposed BOLDSimNet model measures the similarity between MTE-based brain networks for two cognitive tasks. Second, the model proposes a functional imaging-based approach to handle differences in brain network size by substituting, inserting, or removing nodes based on functionally similar ROI groups. This method ensures better alignment of brain networks while minimizing graph transformation loss.

\section{Methods}
\subsection{Data and Preprocessing}
We analyzed the fMRI data of 40 participants collected from the CMI-HBN dataset \cite{CMI-HBN}. During the fMRI scan, participants alternated between task and rest sessions. In the task session, they performed the predictive eye estimation regression (PEER) task, focusing on a point that changed every four seconds. Participants were divided into a child group (CHD; 6--10 years) and an adolescent group (ADO; 11--17 years) for analyses. The ts-fMRI and rs-fMRI data were collected in two runs of 135 and 375 time points, respectively, with a repetition time (TR) of 0.8 seconds. The total scan duration was approximately 14 minutes. Preprocessing was conducted using FS-FAST from FreeSurfer \cite{freesurfer}, including motion correction, brain extraction, spatial smoothing, slice-timing correction, and intensity normalization. Subsequently, FNIRT (FMRIB’s Nonlinear Image Registration Tool) of FSL \cite{fsl} was applied to register the fMRI voxels to the MNI152 standard space, serving as the reference space for the subsequent atlas-based analysis. For functional parcellation, we used the Yeo 17 network atlas \cite{yeo2011} to subdivide each brain into distinct ROIs and then averaged the BOLD signals within each ROI. Because the Yeo 17 atlas is a finer subdivision of the Yeo 7 atlas, we then grouped functionally similar ROIs under Yeo 7 labels to form higher-level network groupings Fig.~\ref{fig:2}.

\begin{comment}
\begin{table}[!h]
    \centering
    \caption{Group of Yeo 17 Network ROIs by Yeo 7 Network Labels}
    \resizebox{\textwidth}{!}{
        \begin{tabular}{l l}
            \toprule
            \textbf{Yeo 7-network ROIs} & \textbf{Yeo 17-network ROIs} \\  
            \midrule
            Visual & Visual A, Visual B \\
            Somatomotor & Somatomotor A, Somatomotor B \\
            Dorsal Attention & Dorsal Attention A, Dorsal Attention B \\
            Salience/Ventral attention & Salience/Ventral attention A, Salience/Ventral attention B \\
            Limbic & Limbic A, Limbic B \\
            Control & Control A, Control B, Control C \\
            Default mode & Default mode A, Default mode B, Default mode C, Temporal Parietal \\
            \bottomrule
        \end{tabular}
    }
    \label{tab:basic_table}
\end{table}
\end{comment}

\begin{figure}[!t]
    \centering
    \includegraphics[page=2, width=1\textwidth]{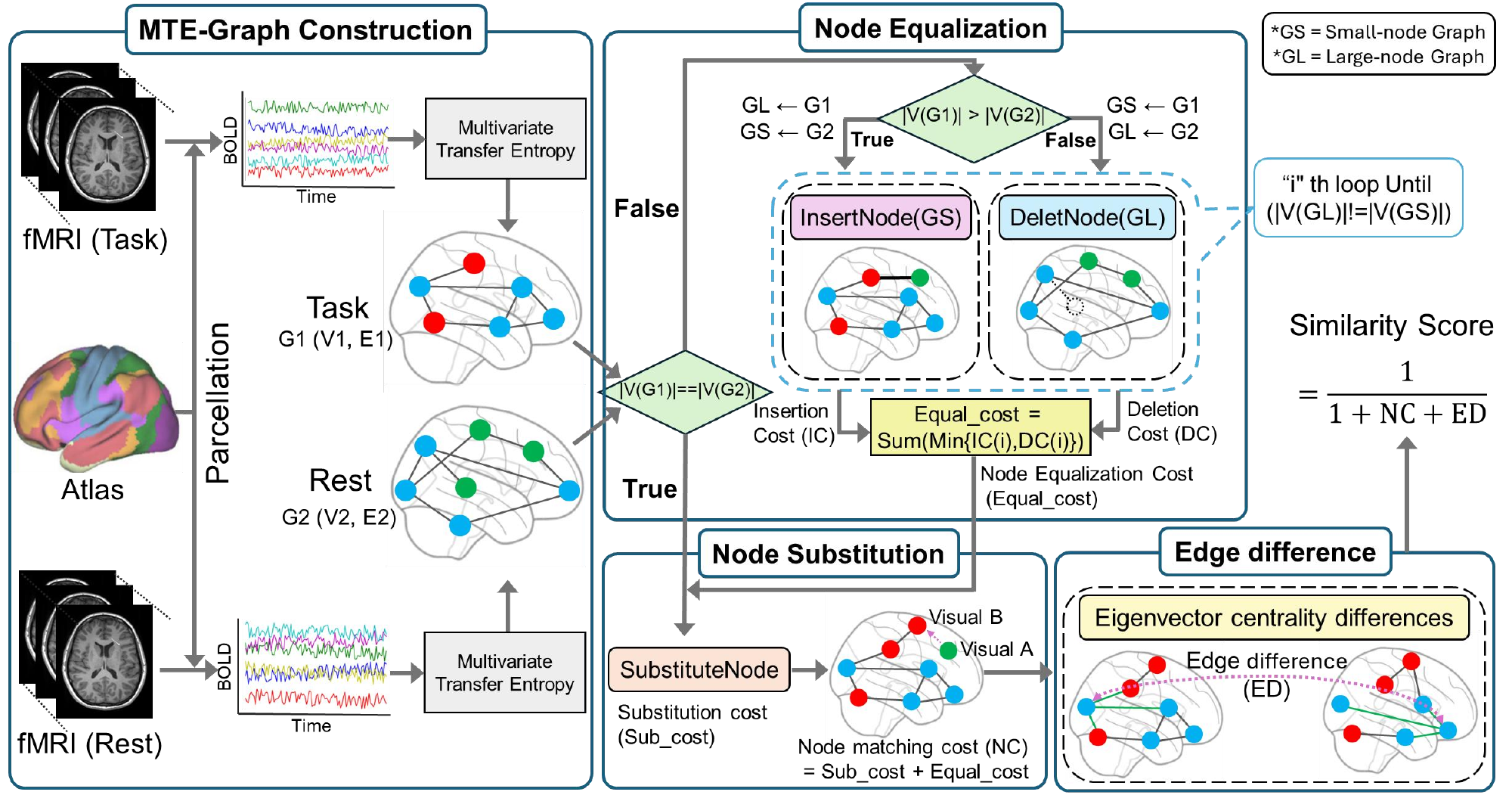}
    \caption{BOLDSimNet calculates the node matching cost and edge difference using set of algorithms to find the similarity score between task and resting state fMRI.}
    %BOLDSimNet: Calculate the similarity between task and resting-state fMRI using algorithms to measure node matching cost and edge difference, and compute the final similarity score.}
    \label{fig:1}
\end{figure}

\subsection{MTE-Based Functional Brain Networks}
The MTE-based functional brain connectivity network is constructed using fMRI data with ROIs defined by the Yeo 17 network atlas (mentioned in Fig.~\ref{fig:2}). We define three processes \(X, Y,\) and \(Z\) to represent the source, target, and conditional variables, respectively, in our time series data. Let \(\{X_t, Y_t, Z_t\}\) be the random variables sampled at time \(t\), and \(\{X_t^-, Y_t^-, Z_t^-\}\) denote the corresponding past observations (e.g., \(\{X_{t}, X_{t-1}, \dots\}\)). Then, the MTE from X to Y conditioned on Z is defined as follows (\ref{eq:MTE}) \cite{montalto2014mute,das2024simbrainnet,chakladar2024brain}:

\begin{equation}
MTE_{X \to Y \mid Z} 
= \sum p\bigl(Y_{t+1}, Y_{t}^-, X_{t}^-, Z_{t}^-\bigr)
\, \log \!
\biggl(
 \frac{p\bigl(Y_{t+1} \mid Y_{t}^-, X_{t}^-, Z_{t}^- \bigr)}
      {p\bigl(Y_{t+1} \mid Y_{t}^-, Z_{t}^- \bigr)}
\biggr)
\label{eq:MTE}
\end{equation}

where, \(p(Y_{t+1}, Y_{t}^-, X_{t}^-, Z_{t}^-)\) is the joint probability distribution of the current and past states, and \(p(Y_{t+1} \mid Y_{t}^-, X_{t}^-, Z_{t}^- \bigr)\) is the conditional probability of $Y_{t+1}$ given the past values of $Y_{t}^-$, $X_{t}^-$, $Z_{t}^-$. We compute \(MTE_{X \to Y \mid Z}\) for each pair of ROIs to form a directed adjacency matrix, thereby constructing the MTE-based causal brain connectivity network.

%%%

% Algorithm 1
\begin{algorithm}[!h]
\caption{BOLDSimNet}
\label{alg BOLDSimNet}
\algsetup{linenosize=\tiny }
\footnotesize
\DontPrintSemicolon 
\KwIn{MTE-based fMRI Brain networks $G_1(V, E)$ and $G_2(V, E)$}
\KwOut{Similarity score}
$NC \gets 0$, $ED \gets 0$ \;
Dictionary $D_{fs} \gets \{\{Yeo\ 17\ ROIs\}: Yeo\ 7\ ROIs\}$ \;
Dictionary $D_{imp} \gets \{imp\_nodes, \{edge,edge.value\}\}$ \;

\If{$V(G_1) = V(G_2)$}{
    $NC \gets$ SubstituteNode $(G_1, G_2, D_{fs})$ \;
    $ED \gets \sum \bigl| \text{Eigenvector\_central}(G_1) - \text{Eigenvector\_central}(G_2) \bigr| \; $
}
\Else{
    $G_S \gets \text{graph with fewer nodes between } G_1 \text{ and } G_2$\;
    $G_L \gets \text{graph with more nodes between } G_1 \text{ and } G_2$\;
    \While{$|V(G_L) - V(G_S)| \neq 0$}{
        $Delet\_cost \gets$ DeleteNode $(G_L, D_{fs})$ \;
        $Insert\_cost \gets$ InsertNode $(G_S, D_{imp})$ \;
        \uIf{($Delet\_cost > Insert\_cost$)}{
            $G_S \gets$ Updated Graph after insertion \;
            $Equal\_cost.add(Insert\_cost)$ \;
        }
        \Else{
            $G_L \gets$ Updated Graph after deletion \;
            $Equal\_cost.add(Delet\_cost)$ \;
        }
    }
    $Sub\_cost \gets$ SubstituteNode $(G_S, G_L, D_{fs})$ \;
    $NC \gets Sub\_cost + \sum {Equal\_cost}$ \;
    $ED \gets \sum \bigl| \text{Eigenvector\_central}(G_S) - \text{Eigenvector\_central}(G_L) \bigr| \; $
}
$Similarity\ score \gets \frac{1}{1+NC+ED}$ \;
\Return $Similarity\ score$\;
\end{algorithm}

\subsection{Measuring Functional Brain Network Similarity}

% Algorithm 2
\begin{algorithm}[!h] %% Dont change the [!t] command
\caption{SubstituteNode}
\label{alg:SubstitutionNode}
\algsetup{linenosize=\tiny}
\footnotesize
\DontPrintSemicolon 
\KwIn{MTE-based fMRI Brain networks $G_1, G_2$ and Dictionary $D_{fs}$}
\KwOut{Updated graphs $G_1$ and $Sub\_cost$}

$nonMatching\_nodes \gets V(G_1) - (V(G_1) \cap V(G_2))$\;
$Sub\_cost \gets 0$\;

\ForEach{$(n \in nonMatching\_nodes) \ \&\ (k \in V(G_2))$}{
    \uIf{$any(D_{fs}[n] == D_{fs}[k])$}{
        $sub\_node \gets \text{any}(k) \text{ where } k \in D_{fs}[n]$\;
    }
    \Else{
        $sub\_node \gets Max(degree(k),\ k \in V(G_2))$\;
    }
    $sub\_edges \gets \sum_{\text{edge} \in G_2.edges(sub\_node)} G_2.MTE(edge)$\;
    $n\_edges \gets \sum_{\text{edge} \in G_1.edges(n)} G_1.MTE(edge)$\;
    $Sub\_cost += |sub\_edges.value - n\_edges.value|$\;
}
%$Updated\ G_1 \gets G_1\{(V \in sub\_node), (E \in sub\_edges)\}$\;
$G_1 \gets G_1\Bigl(\,(V \setminus \{n\}) \cup \{sub\_node\},\;(E \setminus \{edges(n)\}) \cup \{edges(sub\_node)\}\Bigr)$\;

\Return $Sub\_cost$\;
\end{algorithm}

The proposed Algorithm BOLDSimNet (Fig.~\ref{fig:1}, Algorithm 1) consists of three subalgorithms: SubstituteNode, DeleteNode, and InsertNode (Algorithm 2--4).

\begin{figure}[!t]
    \centering
    \includegraphics[page=2, width=1.0\textwidth]{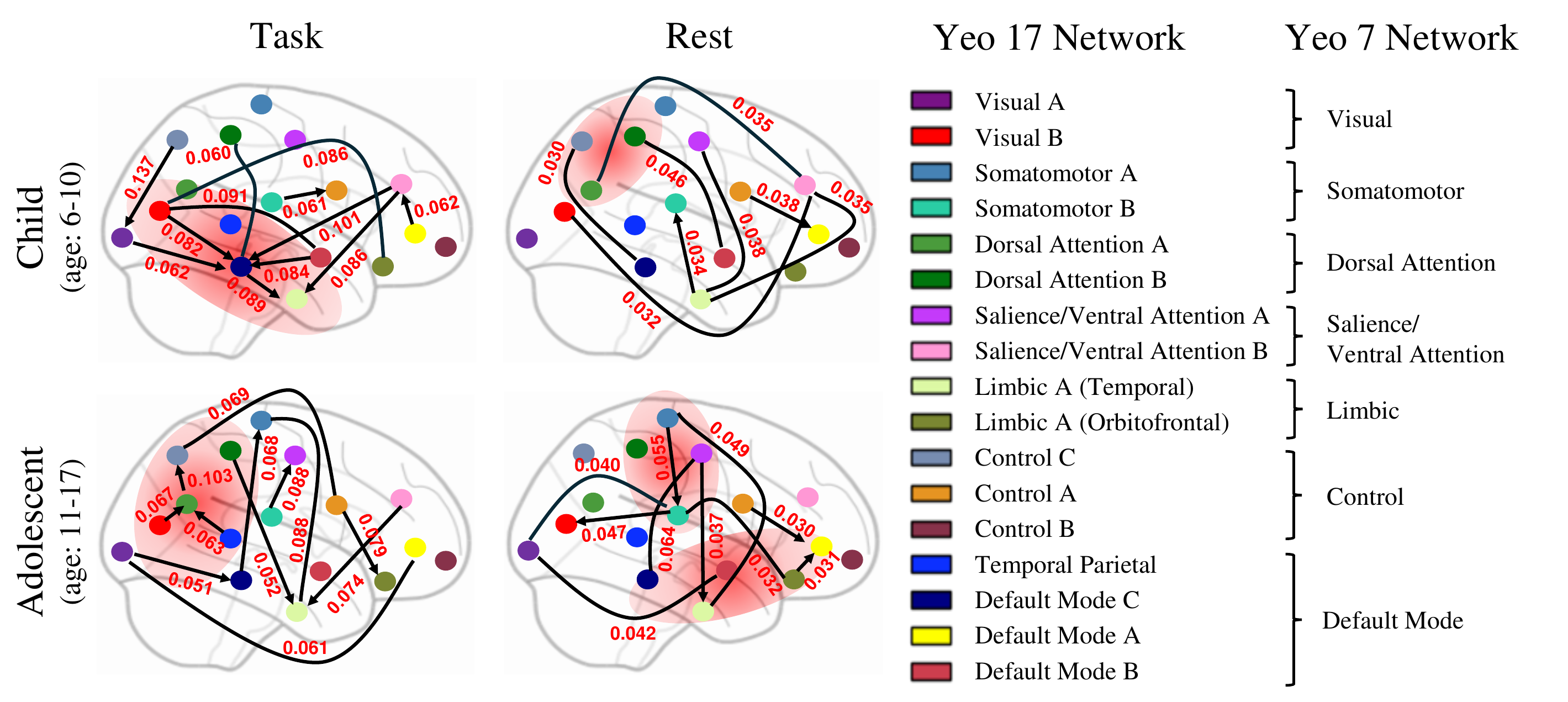}
    \caption{Brain networks evaluated using MTE during the task and rest. The directed edge values represent averaged MTE within each age group. The method effectively demonstrates greater activation in adolescents than children in DAN during task and in DMN during rest.
    }
    \label{fig:2}
\end{figure}

\textbf{BOLDSimNet}: Using MTE, fMRI data from different cognitive states in task and resting state are converted into corresponding brain networks, denoted as \( G_1(V, E) \) and \( G_2(V, E) \). In the graph \( G_1 \) and \( G_2 \), fMRI ROIs are vertices (\( V \)), and edges (\( E \)) represent MTE values between them. Dictionary $D_{fs}$ comprises higher-level groupings of functionally similar ROIs from the Yeo 17 atlas (Fig.~\ref{fig:2}). We use this dictionary to substitute or delete nodes with similar functional roles. The MTE p-value was increased from 0.01 to 0.05 to encompass a broader range of important edges/nodes, which are then stored in $D_{imp}$ and used during the InsertNode process. If $G_1$ and $G_2$ have different numbers of nodes, the insertion cost in the smaller graph ($G_S$) and the deletion cost in the larger graph ($G_L$) are compared.By repeatedly choosing and applying the lower-cost operation until both graphs have the same number of nodes, we sum these costs, resulting in the node equalization cost (\( Equal\_cost \)). After the transformation, if the number of nodes is equal, $SubstituteNode$ is invoked to match the nodes in both graphs and compute the substitution cost (\( Sub\_cost \)). The node matching cost (\( NC \)) is then the sum of \( Equal\_cost \) and \( Sub\_cost \). After matching nodes between the graphs, we calculate eigenvector centrality differences (\( ED \)) to measure changes in hub influence. Because eigenvector centrality accounts for both a node’s direct connections and the importance of its neighbors, it captures local and global connectivity shifts across different cognitive states \cite{van2013network}. The final Similarity Score is calculated as the reciprocal of ${1+NC+ED}$.

\textbf{SubstituteNode:} 
In this algorithm, we initially identify the non-matching nodes between $G_1$ and $G_2$. For each such non-matching node (\( n \)) in $G_1$, if a functionally similar node exists within the same \( D_{fs} \) for $G_2$, then we select any of such substitute node (\( sub\_node \)) from $G_2$ and add its edges as replaced edges. On the other hand, if no such similar node exists, the node with the highest degree in \( G_2 \) (\( sub\_node \)) is selected as the replacement.  
% In this algorithm, we identify unique nodes in the first network (\( G_1 \)) that do not exist in \( G_2 \). For each such unique node (\( n \)), if a functionally similar node exists within the same \( D_{fs} \), it is included in \( k.list \). For each \( k \) in \( k.list \), the difference between the total edge weight of \( k \) in \( G_2 \) and that of \( n \) in \( G_1 \) is computed. The node \( k \) with the smallest substitution cost is then selected as the replacement node. In another case, to preserve overall connectivity while substituting nodes, the node with the highest degree in \( G_2 \) (\( imp\_node \)) is selected as the replacement. 
The substitution cost is computed by the absolute difference in edge values between the \( sub\_node \) and \( n \). This process is repeated for all unique nodes until \( G_1 \) and \( G_2 \) become graphs with identical nodes.

% Algorithm 3
\begin{algorithm}[!h]
\caption{DeleteNode}
\label{alg:DeletionNode}
\algsetup{linenosize=\tiny}
\
\DontPrintSemicolon 
\KwIn{MTE-based fMRI Brain networks $G_L(V, E)$, Dictionary $D_{fs}$}
\KwOut{Updated graphs $G_L$ and $Delet\_cost$}

\textbf{Dictionary} $k\_list : \{\text{node, edge\_val}\} \gets NULL$\;
$Delet\_cost \gets 0$\;
\ForEach{$n \in V(G_L)$}{
    \If{$any(D_{fs}[n] == D_{fs}[k])$}{
        \ForEach{$((k \in V(G_L)),\ Edge(n, k) \neq NULL)$}{
            $edge\_val(k) \gets \sum_{\text{edge} \in G_L.edges(k)} G_L.MTE(edge)$\;
            $k\_list.add(k, edge\_val(k))$\;
        }
        $Min\_edge \gets Min(k\_list[edge\_val])$\;
        $Delet\_cost \gets Min\_edge.value$\;
        $fs\_node \gets k\_list[Min\_edge]$\;
        % $V' = (V \setminus \{n\}) \cup \{fs\_node\}$\;
        % $E' = (E \setminus \{edges(n)\}) \cup \{edges(fs\_node)\}$\;
        $G_L \gets G_L\Bigl(\,V \setminus \{n\}, E \cup(\{edges(n)\} \cup \{edges(fs\_node)\})\Bigr)$\;
        
    }
    \Else{
        $del\_node \gets  Min(\text{degree}(n),\ n \in V(G_L
        ))$\;
        $del\_edges \gets \sum_{\text{edge} \in G_L.edges(del\_node)} G_L.MTE(edge)$\;
        $Delet\_cost \gets del\_edges.value$\;
        % $V' = (V \setminus \{n\}) \cup \{del\_node\}$\;
        % $E' = (E \setminus \{edges(n)\}) \cup \{G_L\_del\_edges\}$\;
        $G_L \gets G_L\Bigl(\,V \setminus \{del\_node\}, E\setminus \{del\_edges\}\Bigr)$\;
    }
}
\Return $Delet\_cost$\;
\end{algorithm}

% Algorithm 4
\begin{algorithm}[!h]
\caption{InsertNode}
\label{alg:InsertNode}
\algsetup{linenosize=\tiny}
\
\DontPrintSemicolon 
\KwIn{MTE-based fMRI Brain networks $G_S(V, E)$, Dictionary $D_{imp}$}
\KwOut{Updated graphs $G_S$ and $Insert\_cost$}

$Insert\_cost \gets 0$\;

\ForEach{$(n \in D_{imp}[\text{imp\_nodes}]) \ \& \ (k \in V(G_S))$}{
    \If{$\exists D_{imp}[n.\text{edge}] = (n, k)$}{
        $edge\_val(n) \gets \sum_{\text{edge} \in D_{imp}[n.\text{edge}]} D_{imp}[n.\text{edge.value}](edge)$\;
        $Insert\_cost \gets edge\_val(n)$\;
    }
}
$G_S \gets G_S\Bigl(V \cup \{n\}, E \cup \{n.edges\}\Bigr)$\;

\Return $Insert\_cost$\;
\end{algorithm}

\textbf{DeleteNode:} This Algorithm operates on the large graph \( G_L \), and functional similar ROIs dictionary \( D_{fs} \). We define a dictionary \( k\_list \) to store the edge values of functionally similar nodes(\( fs\_node \)). For each node \(n\) in \(G_L\), we identify the \( fs\_node \) with the lowest edge value, designate that value as the \(Delet\_cost\), and update the graph by reattaching the deleted node’s edges to the \( fs\_node \). If no functionally similar node exists, we remove the node with the lowest degree (for minimum information flow) from \(G_L\), compute its total edge value as the deletion cost, and update the graph accordingly.

\textbf{InsertNode:} This algorithm adds important nodes to the small graph \( G_S \) by using a predefined dictionary \( D_{imp} \). We compute the insertion cost by summing the weights of all edges connecting the important nodes \( n \) (which are not present in \( G_S \)) to the existing nodes \( k \) in \( G_S \). We then update \( G_S \) by inserting the important nodes with their associated edges.

%%%

%\caption{For both groups (Child and Adolescent), ROI-based brain network edges represent the average MTE between source and target nodes.

\section{Result}
\subsection{Developmental Differences in Brain Networks}
This section explores the differences in MTE-based causal brain networks in CHD and ADO under task and resting states. Fig.~\ref{fig:2} illustrates the average MTE connectivity across all ROIs for both groups. During the task, CHD showed heightened DMN connectivity and reduced SMN and DAN activation. In contrast, ADO suppressed the DMN and activated the DAN and Control networks, consistent with previous studies examining age-related tsfMRI activation patterns \cite{cortese2012toward}. During the resting state, CHD exhibited lower DMN and active DAN, whereas ADO displayed heightened DMN and deactivated DAN. This aligns with the known DMN–DAN anti-correlation \cite{fox2005human}, where DMN is typically active at resting state and DAN during task \cite{seitzman2019state}. However, in CHD, the DAN that was active during the task still remained active during rest, indicating that the network did not fully switch to the resting state.

\begin{figure}[!t]
    \centering
    \includegraphics[page=1, width=1.0\textwidth]{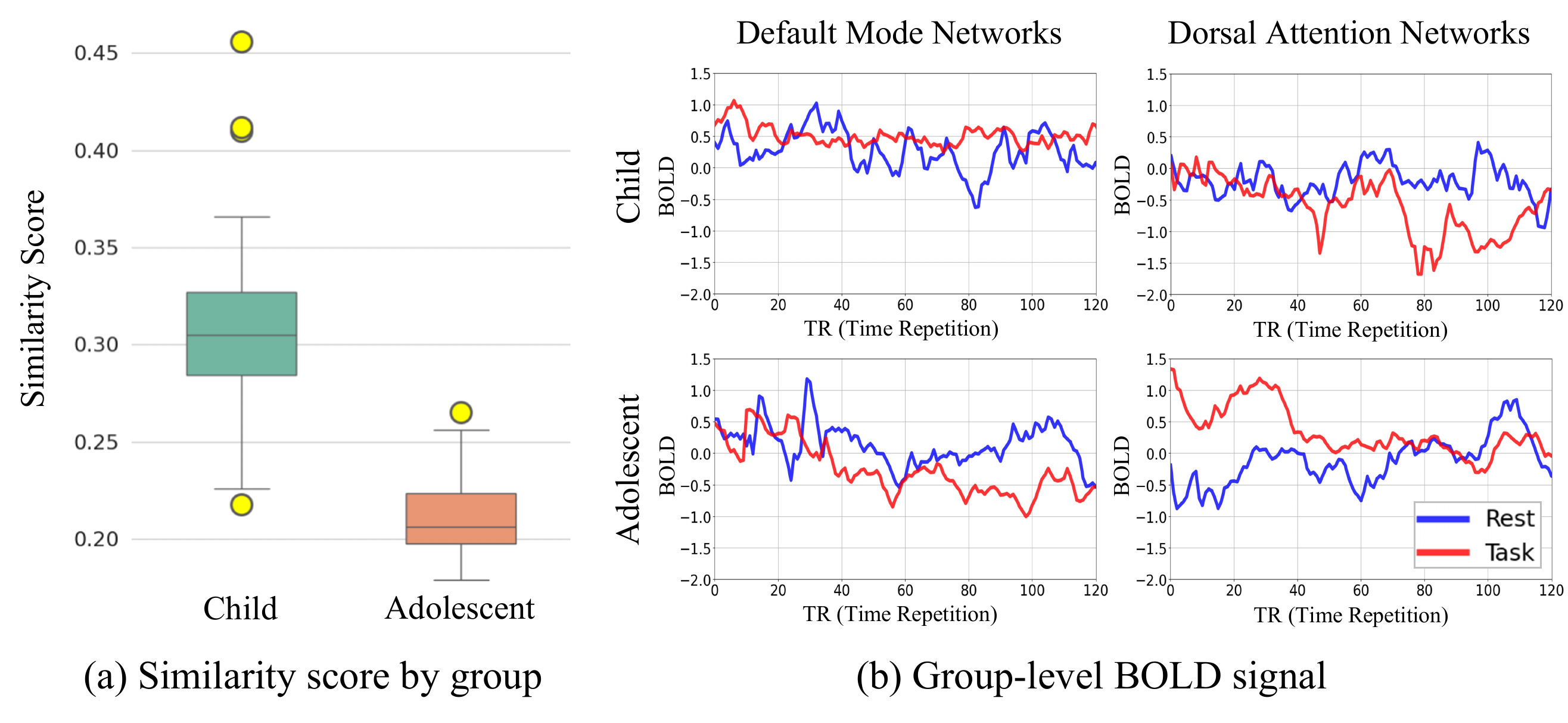}
    \caption{(a) BOLDSimNet similarity score showing brain network similarity between the task and resting states. (b) Mean BOLD signal for DMN and DAN in task and resting states for each age group. Adolescents suppressed DMN and activated DAN when switched from resting to task, whereas children did not. DMN is the average of Default Mode A, B, C, and Temporal Parietal. DAN is the average of Dorsal Attention A and B.
    }
    \label{fig:3}
\end{figure}
% (a) Box plot of similarity scores for all subjects, derived from MTE-based brain networks in task and resting states. (b) Mean BOLD signals for DMN (average of Default Mode A–C, Temporal Parietal) and DAN (average of Dorsal Attention A, B) by group. CHD shows minimal DMN changes and higher DAN activity at rest, whereas ADO exhibits stronger DMN at rest and increased DAN during task, reflecting clearer attentional shifts and lower similarity scores.

\subsection{Analysis of Similarity Scores from BOLDSimNet}
% We evaluate the similarity scores among MTE-based causal connectivity brain networks in both task and resting states. The results shown in Fig.~\ref{fig:3}(a) indicate that subjects in the G1 (Child) exhibit higher similarity scores on average, suggesting that the graphs remain consistent between the two states. This implies that the brain networks undergo less variability during task and rest, reflecting relatively smaller fluctuations in attentional focus and less pronounced switching across states. Conversely, subjects G2 (adolescent) exhibit lower similarity scores, indicating larger attentional fluctuations between the two states and more pronounced network switching. When switching among brain networks is disrupted, individuals may experience diminished attentional capacity and difficulty maintaining focus \cite{fair2010atypical,sridharan2008critical,dibbets2010differential}. 
We evaluate similarity scores among MTE-based causal connectivity networks in task and resting states. As shown in Fig.~\ref{fig:3}(a), CHD exhibits higher similarity scores, indicating stable brain networks with fewer attentional fluctuations and state transitions. In contrast, ADO shows lower similarity scores, suggesting greater attentional variability and more pronounced network switching. Disruptions in brain network switching may lead to reduced attentional capacity and difficulty maintaining focus \cite{fair2010atypical,sridharan2008critical,dibbets2010differential}.
In Fig.~\ref{fig:3}(b), we analyze the mean BOLD signals for the DMN and DAN across all subjects, separated by group. In CHD, the DMN showed no clear difference between the task and the resting state, with relatively higher activation during the task state, whereas the DAN was more active during the resting state. By contrast, in ADO, the DMN exhibited stronger activation during the resting state with a more pronounced gap, while the DAN showed increased activity during the task state. Analyzing BOLD signals in the DMN and DAN effectively identifies differences in attentional focus \cite{dibbets2010differential,metin2015dysfunctional}. These findings support the link between similarity scores and attentional focus in our study.
%In CHD, the DMN showed no clear difference between the task and resting states and notably displayed higher activation during the task, while the DAN was more active at rest. By contrast, in ADO, the DMN displayed stronger activation at rest with a more pronounced gap, while the DAN was more active during the task. Analyzing BOLD signals in the DMN and DAN effectively identifies differences in attentional focus \cite{dibbets2010differential,metin2015dysfunctional}. 
\par
We performed permutation and bootstrap analyses on the similarity scores from CHD and ADO to assess the statistical significance and robustness of their differences. The observed mean difference (Child – Adolescent) was 0.101, which was significant in the permutation test (p = .0084). The bootstrap analysis yielded a 95\% confidence interval from 0.047 to 0.176.

\section{Conclusion}
We propose BOLDSimNet, a framework to assess the similarity between two brain causal connectivity networks using fMRI data collected during task and resting states. By leveraging functionally similar ROIs groups, BOLDSimNet calculates this similarity while minimizing graph transformation loss. We observed that the child group exhibited higher similarity scores than the adolescent group. Using the similarity scores derived from BOLDSimNet, it is possible to detect differences in brain network reconfiguration and changes in attentional focus across different states. Future work will evaluate the proposed model using a larger cohort and a broader range of states.
%In this study, we conducted experiments on a small sample of 40 subjects across PEER task and resting state. 

% Acknowledgments & Disclosure of Interest sections not needed!
% To avoid an accidental anonymity breach, you are not required to include the acknowledgments and the Disclosure of Interest sections for the initial submission phase.  If your paper is accepted, you must reinsert these sections to your final paper. 

\bibliographystyle{unsrt}
\bibliography{ref}

\end{document}